# Evaluating the performance of geographical locations in scientific networks with an aggregation – randomization – re-sampling approach (ARR)

Stefan HENNEMANN, Justus-Liebig-University Giessen

Dept. of Economic Geography, Senckenbergstrasse 1, D-35390 Giessen, Germany, stefan.hennemann@geogr.uni-giessen.de

*Abstract*

Knowledge creation and dissemination in science and technology systems is perceived as a prerequisite for socio-economic development. The efficiency of creating new knowledge is considered to have a geographical component, i.e. some regions are more capable in scientific knowledge production than others. This article shows a method to use a network representation of scientific interaction to assess the relative efficiency of regions with diverse boundaries in channeling knowledge through a science system. In a first step, a weighted aggregate of the betweenness centrality is produced from empirical data (*aggregation*). The subsequent randomization of this empirical network produces the necessary Null-model for significance testing and normalization (*randomization*). This step is repeated to yield higher confidence about the results (*re-sampling*). The results are robust estimates for the relative regional efficiency to broker knowledge, which is discussed along with cross-sectional and longitudinal empirical examples.

The network representation acts as a straight-forward metaphor of conceptual ideas from economic geography and neighboring disciplines. However, the procedure is not limited to centrality measures, nor is it limited to spatial aggregates. Therefore, it offers a wide range of application for scientometrics and beyond.

*Introduction*

The geographical distribution of economic and scientific activity is uneven across nations, regions, and cities. It is in the interest of researchers, politicians, and evaluators to assess and measure the performance of individuals, groups, organization, but also of geographically bounded activity. Research performance is increasingly measured with sophisticated bibliometric methods and the spatial scale is more and more in the focus, opening a new field of *spatial scientometrics* (Frenken et al. 2009).

Performance measurement is in the natural interest of all funding agents that usually act on a national level and of supra-national funding such as in the case of the European Union. But not only funding bodies are interested in getting reliable information about the performance of science systems. As scientific research is a prerequisite for socio-economic development and welfare, all related information tools that constantly lack of comparable and normalized data on the science system output will be interested. One of the problems for the socio-economic information processing is the territorial scale that varies with the availability of other data. Hence, a flexible territorial grouping would be favorable.

To build networks from science system information has several advantages and helps to overcome the necessity of quality assessments of contributions (e.g. number of citations etc.). Instead, network flow analogies help to evaluate the systemic relevance of an actor for the network functioning, because network flows are at the conceptual center of many theories of territorial systems research (e.g. global cities, innovation systems, agglomeration theory). Through geo-referencing/attributing and aggregation all individual agents can be combined to larger groups that may represent research-teams, whole organizations, or result in a projection of different territorial scales (e.g. cities, counties, provinces, states, countries).

The results can be used for information purposes in the sense of a strategic information tool for regional politicians (cf. Brandt 2009), but also to produce robust output indicators of science systems that may be regressed with other economic indicators.

This article aims at contributing to the recent discussion in spatial scientometrics and proposes a simple and reproducible step-by-step introduction of a method to test the significance of spatial network structures at flexible scales. With the approach presented here, the strength of network analytical methods will be combined with a bootstrapping approach that allows for spatial and temporal comparisons.

The article is structured as follows: after this introduction, a brief review of concepts and methods that are related to spatial networks and their evaluation will be given. The subsequent section will unfold the idea of regional betweenness by using a simple example network and explaining the steps that are necessary to deliver a comparable measure of spatial performance. It will be followed by a section that uses empirical bibliographic data to test the proposed approach. Subsequent to this empirical section, the method will be tested with a focus on the reproducibility of the results. A final section will discuss further research suggestions with respect to the method and different data.

*Conceptualizing spatially bounded network performance*

Competitiveness nowadays depends largely on the ability of individuals and organizations to produce, to absorb, to reassemble and to disseminate knowledge (Gertler 2003, 76) and networks are the most likely mode of governance in knowledge economies (de Man 2008). These socio-economic processes and dynamics are usually interfering with the location-related aspects (e. g. institutions). This will lead towards a differentiated activity and performance of places. The economic competitiveness of nations and the contribution of smaller spatial units such as metropolitan areas or cities to this success are evaluated mainly through secondary data from public sources. On the one hand, however, the effects that networks exert on the geographical space are hard to determine through such aggregate data without any relational information. On the other hand, relational data is hard to obtain, and if available, the question is on how to organize the relations, how to aggregate, how to assess the performance, and how to present the results to the ranking avid (scientific) public?

Today, places are seen as the main catalytic space where network flows manifest in performance gains of efficiently acting organizations (cf. Castells 1996). Not the activity in an agglomeration is of central interest, but the capacity to channel flows in a higher order system such as national or global systems. However, places are not actors, but people and organizations valorize places with their relation to other actors in other places. Space and networks are inevitably and reciprocally intertwined: the performance of actors in networks influence the regional perspective of growth and investment in the long run. Regions, in turn, constitute the seedbed for attracting networking actors from outside the region, building bridges to other places with their existing relations. Non-spatial networks are different from structure and behavior, when compared to networks in which actors possess a rigid location (Warren et al. 2002, Rozenfeld et al.

2011). This fact can be capitalized in improving the regional capacity to grow in bridging "geographic holes" (Bell and Zaheer 2007), i.e. occupying scarce positions in the spatial networks, similarly conceptualized by Burt (1995) for spanning strategic social network positions. Long-term growth dynamics can be influenced and prevent from spatially induced lock-ins through regional bounded thinking. Likewise, places with little capacities can channel their activity towards effective positions (cf. Liefner and Hennemann 2011). Thus, there are different positions in networks to be occupied and the actors can differ in their ability to integrate themselves into the existing network structures. This will affect the efficiency and the effectiveness of their positions, i.e. the overall ability to utilize scarce resources to create, absorb, and broker knowledge.

Essentially, network approaches possess the analytic capacity to serve as strategic regional information tool for politicians (Brandt et al. 2009). This view acknowledges that networks are important vehicles for the exchange of information, even for the local scale and that knowledge creation in scientific networks has its importance for the overall socio-economic development.

The growing number of articles on the performance measurement in scientific systems suggests an increasing awareness of researchers on regional differences in scientific efficiency. Recently, Bornmann & Leydesdorff (2011) proposed a method for the visual representation of over- and under-performing cities with respect to scientific activity, however, without using an explicit networking perspective. Stock (2011) uses the idea of the *informational city*, which comes close to Castell's original notion, to review and conceptualize the fundamental necessity of knowledge infrastructure and networks for the effective scientific knowledge creation. With changing environments and more an more knowledge-based industries, intensive interaction between different actors such as governments, universities, and companies are increasingly dependent on hard and soft location factors that facilitate (scientific) knowledge creation (cf. Potter & Watts 2010).

Other examples use a more direct network perspective. Recently, Hennemann et al. (2011a) found that scientific collaboration in epistemic communities is mainly a localized activity in many scientific fields, i.e. the probability of collaboration for spatially adjacent located researchers is an order of a magnitude higher than for collaboration that involves researchers from different countries. Pepe (2011) delivers one explanation for this phenomenon with presenting empirical evidence for the overlap of personal acquaintanceships and professional collaboration in science.

Obviously, the cohesiveness of local science clusters, based on socio(-economic) ties is a very relevant feature for creating and disseminating state-of-the-art scientific knowledge - an idea that is supported by the work of Leydesdorff & Rafols (2011) on the spatial diffusion of new technologies. Related to this knowledge flow analogy, Yan & Sugimoto (2011) found that various types of distances (e.g. geographic, cognitive, social) play a significant and differentiating role in citing another researcher's work in the sense that citation tend to be a localized phenomenon.

Most of these examples use bibliographic data such as scientific articles or patents, simply because this information is available and comprehensive, compared to primary surveys. However, there are some important prerequisites for the calculation and methodological constraints that need to be accounted for, but that are frequently disregarded. The calculation need to be valid, i.e. is the combination of data, indicators and calculation method, able to represent the phenomenon (e.g. indicate for the scientific capacity of regions)? Another requirement is related to the reliability of the results, i.e. can the results be reproduced, and, more importantly, can results of different analyses be compared to one another and over time?

*Current approaches to measuring spatially bounded networking activity*

The Globalization and World Cities Research (GaWC) Interlocking Network Model approach was one of the first to comparably measure the networking among metropolises worldwide, based on the interaction of advanced producer services (Taylor 2001, among many others). With this model, the authors combine the territorial perspective with a relational view on the measurement of socio-economic processes. It makes use of the interlocking between company networks that define the connectivity between cities. Basically, the degree centrality is the only indicator in this system of cities. The interlocking network model was criticized for its narrative character, for the understanding of the network concept, and for the way of computation (cf. Nordlund 2004). However, the method is still and increasingly in use, as it is still one of few attempts to analyze city relations instead of resource based assessment of global city capacities. This persistent use of the approach can be interpreted as success or as a desire for methods that enable regional comparisons of networking activity and performance. Over the last ten years, various extensions for the interlocking network model have been proposed to improve the method's ability to describe the global city system organization (cf. Neal 2011). However, the system's perspective is still not fully acknowledged and controlling for network effects is not taking place. This affects the robustness of the results and the reproducibility of results (cf. Neal 2010).

Paci and Usai (2009) use another common way of using network representations for the evaluation of regional performance. They use aggregate patent citations of individual inventors to link regions at each citation of a patent. This leads merely to a descriptive counting of regions citing regions, rather than essentially evaluating the networking performance (not to mention the lack of data normalization). Due to the systemic effects of strong interaction and interdependence of nodes and edges (emergence) most information about the complexity of the system will be lost with such reduced procedure. Further, a reasonable level of aggregation should be explainable by a conceptual homogeneity of the aggregates, i.e. the combined individuals need to be somehow similar/related to one another. This may be assumed for members of the same organization (e. g. university, company; although organization theorists may intervene), but can be seriously questioned for regional aggregates, where individual members of different organizations are lumped together only based on their common location. It is this type of application, which let Newman (2003, 175) conclude that many empirical network studies tend to have major flaws in terms of being a reasonable projection of the underlying interconnections.

In summary, three main problems exist in many studies that employ network representations for regional analyses. First, the networks need to be comprehensive to avoid missing links that create immediate shortcuts (cf. Canter and Graf 2008). Second, an unsuited level of aggregation leads towards a combination of actors who have little in common. These combinations (e.g. all firms in a region) presume fully cross-connected regional actors and loose most of the relational information. Third, scaling and size effects in networks hinder from sound comparisons at the regional level, especially for centrality measures (Bonacich 1991, 159).

All three problems are potentially present in spatial networks, where the complexity of the interactions that produces hardly comparable results. Technically, they represent a model where n parameters (=nodes) have to be estimated. However, there is a tradeoff to be made between the accurateness of the network projection and data-organization along with calculation effort. The following procedure is accounting for all three issues and opening a new perspective on the measurement of spatial networking performance and the capacity to channel knowledge flows. It consists of three basic steps: first, calculation of the node-based empirical network parameter, second, the aggregation step, and third, the normalization and bootstrapping for significance testing and parameter estimation.

*The aggregation – randomization – re-sampling approach (ARR)*

*Flow centrality measures and group performance*

Grouped aggregates of network measures as proposed by Wasserman and Faust (1994, 191) seek to assess parts of a network with respect to their heterogeneity of the group members. In addition, they are suited to evaluate the average capacity to constitute a central group of the network. There are several centrality concepts in network science. However, information and knowledge flows in the sense of spatial capacity to participate in and digest of knowledge, are best captured by the betweenness centrality. In the case presented here, all groups are spatial units, therefore we propose the measure of regional betweenness centrality ($C_{RB}$) that acknowledges the networking actor as the basic object of investigation, rather than starting from a spatial aggregated unit (e. g. state, county, city). Anyway, the concept can simply be transformed to other group-centered conceptions such as teams, working groups, or organizations.

The aggregation of node-level measures needs to include differences between the nodes to maintain as much information as possible. A weighting scheme would have to reflect the different abilities to take effective *influence* on the network. The k-core concept, proposed by Kitsak et al. (2010), allows for differences in the potential of a node to infect the whole system. This epidemiologically undesired ability would be an important feature for knowledge networks and the ability of nodes to influence knowledge flows. Thus, weighting using to the k-core number of a node is adjusting the relevance of each node with respect to the overall system positioning. This reflects the different ability of hubs of the same degree to effectively spread information over a network.

K-cores are extracted by iteratively removing nodes, starting with nodes of degree 1 until all nodes are removed. All nodes that had a degree of 1 after any iteration step are given a k-core of 1. If no node of degree 1 is left in the network, all nodes of degree two are removed, and so on. Therefore, k-cores are not only considering the local network position (degree), but identify if this position is stabilizing the network, i.e. if the node survives a long or a short time. For example, node 5 in figure 1 has a degree of 4, but a k-core number of 2 because nodes 10 and 11 are removed in an earlier iteration step, leaving node 5 with only a degree of 2. Batagelj and Zaversnik (2003) provide an efficient algorithm for the k-core calculation.

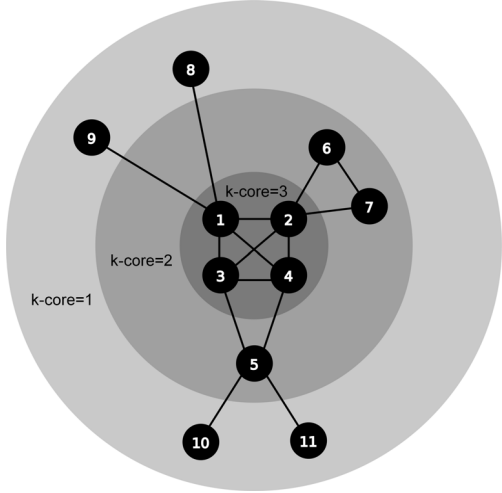

Figure 1: Illustration of the K-Core/Shell Position (adopted and changed from Kitsak et al. 2010)

Finally, $C_{RB}$ is calculated as the k-core-weighted mean of the betweenness centrality over all nodes that are located in the respective spatial unit, scaled between 0 and 1. A value of 0 means that there is no information flow on average through the nodes that are located in the region. At a value of 1, all shortest paths go through all the nodes in the region (see equation 1).

We calculate the regional betweenness of a region *j* with

$$C_{RB}(j) = \sum_{v=0}^{n_j} C_B(v) * \frac{core(v)}{\sum core(v)_j} \quad (1)$$

where, $C_B(v)$ is the betweenness of node *v* and *core(v)* is the k-core value of node *v*.

The aggregation unit is a flexible construction of any group that is larger than an individual node. The empirical examples that are presented below, evaluate the provincial level and the national level performance, simply for reasons of simplicity of the presentation.

*Normalization with Null-models*

Networks are largely *n,k*-dependent, i.e. network properties such as graph-level indices are generally incomparable between networks of different size and density without adjustments (cf. van Wijk et al. 2010, Hennemann et al. 2011b). However, finding sound baseline models (also termed Null-models) of the empirical networks is non-trivial, but essential to control for these network related size-density effects. Hence, the Null-model in a statistical sense should serve as an objective reference model, although there is no *one* agreed reference for networks. Only with such a reference, the graph level and node level features of the empirical network can be compared to a grounded expectation that is based on the congruence of the topology of the empirical and the theoretical network.

There are mainly three ways proposed in the literature to achieve a comparable randomized network: (1) simple parameter corrections, using the parameters of one of the popular network models of the same size and density (e.g. Barábasi-Albert, Watts-Strogatz), (2) exponential random graph models/p*-models that produce multivariate estimates based on Markov chain Monte Carlo algorithms (Morris et al. 2008, and the authors therein), (3) randomization and simulation approaches (similar to bootstrapping techniques). While (1) is producing baseline models that have little in common with the empirical networks (e. g. only the average degree $\langle k \rangle$ is used), approach (2) is highly complex and hard to comprehend. Moreover, p*-models that mimic properties of the underlying empirical network require extremely long computing time. However, it is not in the scope of this article to compare these techniques with one another. Van Wijk et al. (2010) are providing a comprehensive discussion of the mainstream techniques along with robustness testing.

For the purpose of univariate parameter estimation, randomization bootstrapping is a reasonable approach that minimizes effort and maximizes the outcome. The randomization of a given (empirical) network based on edge swapping was first proposed by Maslov and Sneppen (2002). A swap is realized with randomly picking two edges *a-b* and *c-d* from the empirical network and cross-rewire them into *a-c* and *b-d*. To allow for a connected network after swapping, an iterative Markov Chain Simulation Method was proposed by Gkantsidis et al. (2003) to shuffle the network effectively and efficiently and still preserving the original degree distribution as the main constituent of any network (cf. Gkantsidis et al. 2003). The result is a statistically uncorrelated Null-model of the empirical network.

However, this randomized network is just *one* representation of the empirical network. To reach higher confidence about the parameter estimates, a common approach in interference statistics is a re-sampling technique called bootstrapping, especially when the underlying distribution is unknown or highly non-linear (Efron 1979). Due to this simulation-based randomized sampling, this procedure is also referred to as *Monte Carlo Test*. Bootstrapping is used to estimate standard deviations, unknown parameters and confidence intervals, or to calculate a test-statistic for an empirical value. Generally, this procedure is suited to handle complex empirical data, when no asymptotic estimates are available or theoretical distributions are absent (Boos 2003, pp. 117, 168). An advantage of this technique is that the analysts need not to possess comprehensive theoretic knowledge about non-linear approaches and distributions (Davison et al. 2003, p. 142).

A repeated randomization through edge-swapping yields a mean of the parameter that is equivalent to a test value of the Null-model distribution, just like T for the student's t-distribution. Using the standard deviation *SD* of this mean value gives an idea of the error of the parameter estimate (here +/-2 *SD* were used to define a very significant deviation, differences between +/-1 and +/-2 *SD* is interpreted as moderate significant). Thus, the deviation represents the significance of the empirical parameter under the condition of structurally equivalent randomized versions of the network.

Finally, the randomized parameter can be used to normalize the empirical network parameter by dividing the empirical value through the randomized value. This normalized parameter is a proxy for the efficiency of the node. The normalized regional betweenness $p_{rand}$ is calculated with

$$p_{rand} = C_{RB}(v)_{emp} / \langle C_{RB}(v) \rangle_{rand} \qquad (2)$$

and can be compared across nodes of the same network and across different networks (cf. McAuley et al. 2007).

In network sciences, parametric bootstrapping along with randomization based on edge-swapping may be used for all graph level and node level indices, except the degree centrality.

*Technical realization with NetworkX (Python package)*

The calculation has been realized with the NetworkX package for Python (http://networkx.lanl.gov).

1) Read in the empirical network data, containing group attributes for the nodes

2) Calculate the desired empirical network property for the largest component (e.g. $C_{RB}$ using functions `betweenness_centrality()`, `core_number()` with eq. 1)

3) Swap the edges in the largest component of the empirical network with the function `connected_double_edge_swap()` (about 10 times the number of the empirical edges have proven to deliver robust results in terms of true randomized versions of the empirical network)

4) Calculate randomized network properties (e. g. $C_{RB}$ using functions `betweenness_centrality()`, `core_number()` with eq. 1)

5) Repeat steps 3) and 4) for a reasonable number of times (more than 100 configurations have proven to deliver robust and reproducible values)

6) Calculate the test parameter (mean, standard deviation, confidence interval) and the normalized parameter $p_{rand}$

*A simple explanatory example*

Networks are per se placeless, however, the actors are often present in a specific location, which can be modeled by assigning locational attributes to the network nodes. In the following example, there are three regions A, B, and C that host 6, 1, and 6 nodes respectively in situation a). In situation b) region B receives one additional node and the star-like formation in region C evolves into a ring. Situation c) represents a further evolution into a more interlinked network.

The results for network a) show that region C is the least important region in terms of knowledge flows. The hub C8 is evaluated less relevant through a comparably low k-core value, although this node has the highest degree. This effect of marginalization of hubs has been dissected by the k-core weight (cf. Kitsak et al. 2010). The star formation, especially in a peripheral position, is marginalizing the flow capacity of the whole region. The peripheral position is empirically weaker than expected by pure chance, because the relatively high number of edges could have been spent more effectively. This gives an indirect explanation that a local star formation is a relatively inefficient structure for the system flow. Region A is evaluated better than C and receives a score that can be expected from the network structure. Both regions were identified as marginal according to their peripheral position in the system. However, a star formation with a pronounced hub-and-spoke structure is judged least optimal in terms of networks flow capacity, when compared to the ring formations (region A). Region B, which is located in the visual center of network a), receives the highest scores. This result is far from being random, suggesting a very significant position and performance of the node in the region B.

With the changes to network b) region B loses its outstanding position, due to the introduction of a new node to this region that is less favorable connected, thus having a low k-core value. However, the empirical score is still significantly higher than expected. Region A loses in relevance, as indicated by the change of the normalized value $p_{rand}$ that went down from 1.033 to 0.871. The indicated performance for region C goes up from 0.114 to 0.179. This empirical result is confirmed, when allowing for structural changes between both networks: the normalized value indicates a dramatic improvement of the position (0.510 to 0.861).

In network c), region A is still losing ground, now turning into a significant under-performer, whereas region C continues to improve its performance (although still does not occupy a significant position). Region B is again the one that is central in terms of flows in the network, which is also the intuitive positioning of that region, based on a visual assessment.

Generally, the normalized value $p_{rand}$ allows to cross-compare the node performances within and over networks. The difference between the empirical value and the testing value, expressed in standard deviations determines the significance of the empirical value.

The next section applies this procedure to some empirical datasets containing bibliographic information. This data can be understood as source that proxies for the prerequisite of many economic processes: scientific knowledge.

*Empirical network examples*

*Cross-sectional data: global science networks*

The first empirical dataset contains bibliographic data about research on nanotubes, the second set deals with research on the H5N1 virus. Both sets come from Thompson Scientific's SCI-Expanded index and contain all articles that were written between 2004 and 2008 on the respective topic. The networks based on this data represent a co-authorship relation between all involved affiliations of any article, i.e. the nodes in the networks represent an organization rather than an individual author. To aggregate on the organization level eases the process of the data cleaning. The error that may arise from this aggregation is likely to be very small, because the individuals are likely to work in the same department, anyway (or at least they know each other very well). For a detailed description of the network data see Hennemann et al. (2011a). A regionalization on the national level was chosen for the clarity of the presentation here.

These results represent the ability of countries to be a central part of the knowledge flows in nanotube research and H5N1 research. Just for an exemplary interpretation, Ireland ranks first in nanotube research with a significantly higher regional betweenness than expected by chance. However, the performance of Russia is comparably better, since the normalized value $p_{rand}$ is higher than for the Ireland case. It can be hypothesized that all contributing organizations are relatively central and relevant for the knowledge flow in this network. The Swiss' performance is in the expected range, whereas the nanotube research performance in Germany or Belgium is significantly lower than that to be expected from the underlying network structure.

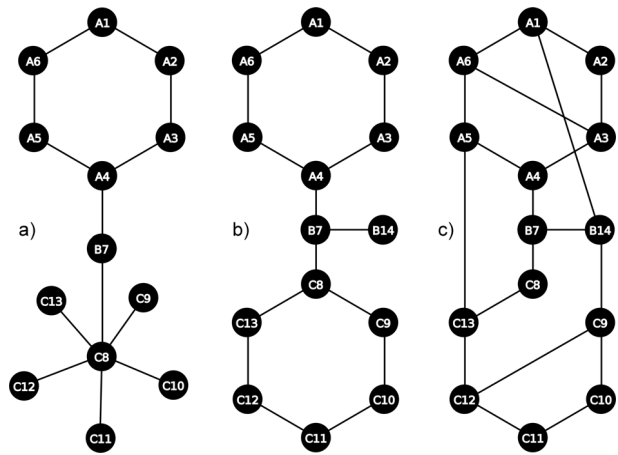

|  | $C_{RB}$ (emp) |  | $C_{RB}$ (rand) | StDev (rand) | $p_{rand}$ |
|---|---|---|---|---|---|
| network a) |  |  |  |  |  |
| Region A | 0.189 |  | 0.183 | 0.042 | 1.033 |
| Region B | 0.545 | ** | 0.172 | 0.131 | 3.170 |
| Region C | 0.114 | ** | 0.223 | 0.019 | 0.510 |
|  |  |  |  |  |  |
| network b) |  |  |  |  |  |
| Region A | 0.179 |  | 0.206 | 0.052 | 0.871 |
| Region B | 0.410 | * | 0.239 | 0.086 | 1.714 |
| Region C | 0.179 |  | 0.209 | 0.055 | 0.861 |
|  |  |  |  |  |  |
| network c) |  |  |  |  |  |
| Region A | 0.106 | * | 0.138 | 0.026 | 0.771 |
| Region B | 0.209 | * | 0.157 | 0.050 | 1.337 |
| Region C | 0.115 |  | 0.104 | 0.023 | 1.098 |

Figure 2: Example Networks with Regional Node Attributes and typical Network Formations (500 configurations, with 100 Edge Swaps at each Step)

*a) Nanotube research*

| # | Country | No. Nodes | $C_{RB}$ (emp) | $C_{RB}$ (rand) | StDev (rand) | $p_{rand}$ |
|---|---|---|---|---|---|---|
| 1 | Ireland | 9 | 0.007 ** | 0.005 | 0.000 | 1.34 |
| 2 | Singapore | 22 | 0.006 ** | 0.004 | 0.000 | 1.38 |
| 3 | England | 87 | 0.004 ** | 0.003 | 0.000 | 1.22 |
| 4 | Japan | 412 | 0.003 ** | 0.003 | 0.000 | 1.10 |
| 5 | Russia | 105 | 0.003 ** | 0.002 | 0.000 | 1.60 |
| 6 | PR China | 574 | 0.003 ** | 0.002 | 0.000 | 1.30 |
| 7 | USA | 675 | 0.003 ** | 0.003 | 0.000 | 1.09 |
| 8 | France | 176 | 0.003 * | 0.003 | 0.000 | 0.96 |
| 9 | Taiwan | 147 | 0.002 ** | 0.001 | 0.000 | 1.56 |
| 10 | Switzerland | 35 | 0.002 | 0.002 | 0.000 | 1.01 |
| 11 | Belgium | 31 | 0.002 ** | 0.002 | 0.000 | 0.84 |
| 12 | South Korea | 221 | 0.002 ** | 0.002 | 0.000 | 0.92 |
| 13 | Ukraine | 18 | 0.002 ** | 0.001 | 0.000 | 1.49 |
| 14 | Denmark | 20 | 0.002 ** | 0.001 | 0.000 | 1.81 |
| 15 | Sweden | 25 | 0.002 | 0.002 | 0.000 | 1.00 |
| 16 | Spain | 75 | 0.002 | 0.002 | 0.000 | 1.01 |
| 17 | Israel | 16 | 0.002 | 0.002 | 0.000 | 1.12 |
| 18 | Australia | 52 | 0.002 ** | 0.001 | 0.000 | 1.25 |
| 19 | Canada | 76 | 0.002 ** | 0.001 | 0.000 | 1.59 |
| 20 | Germany | 168 | 0.002 ** | 0.003 | 0.000 | 0.66 |

*b) H5N1 (bird-flu virus) research*

| # | Country | No. Nodes | $C_{RB}$ (emp) | $C_{RB}$ (rand) | StDev (rand) | $p_{rand}$ |
|---|---|---|---|---|---|---|
| 1 | USA | 225 | 0.011 ** | 0.005 | 0.000 | 1.98 |
| 2 | PR China | 118 | 0.010 ** | 0.006 | 0.000 | 1.58 |
| 3 | Switzerland | 8 | 0.006 | 0.006 | 0.001 | 0.98 |
| 4 | Germany | 34 | 0.006 ** | 0.001 | 0.000 | 3.96 |
| 5 | Netherlands | 30 | 0.006 ** | 0.003 | 0.000 | 1.78 |
| 6 | Thailand | 44 | 0.005 ** | 0.004 | 0.000 | 1.18 |
| 7 | England | 52 | 0.005 ** | 0.004 | 0.000 | 1.47 |
| 8 | Vietnam | 21 | 0.005 ** | 0.009 | 0.001 | 0.59 |
| 9 | Egypt | 4 | 0.005 | 0.006 | 0.001 | 0.84 |
| 10 | Singapore | 16 | 0.004 ** | 0.002 | 0.000 | 2.30 |
| 11 | Japan | 79 | 0.004 * | 0.004 | 0.000 | 0.93 |
| 12 | Austria | 4 | 0.003 | 0.004 | 0.001 | 0.87 |
| 13 | Italy | 21 | 0.003 ** | 0.004 | 0.000 | 0.73 |
| 14 | Taiwan | 19 | 0.003 ** | 0.001 | 0.000 | 4.28 |
| 15 | Australia | 49 | 0.003 * | 0.003 | 0.000 | 1.10 |
| 16 | Turkey | 5 | 0.003 | 0.003 | 0.001 | 0.95 |
| 17 | Nigeria | 13 | 0.003 | 0.003 | 0.000 | 1.06 |
| 18 | Canada | 38 | 0.003 | 0.003 | 0.000 | 1.04 |
| 19 | South Korea | 20 | 0.003 ** | 0.002 | 0.000 | 1.43 |
| 20 | France | 44 | 0.002 ** | 0.002 | 0.000 | 1.53 |

Table 1: Comparison of the Country Performance in a) Nanotube and b) H5N1 (Bird-Flu Virus) Research in the Years between 2004 and 2008 (in descending Order by CRB (emp); Largest Component: a) 3,895 Nodes /15,933 Edges, b) 1,023 Nodes / 3,867 Edges) each 500 Configurations/10x Swaps

In the case of Germany, this under-performance is completely reversed for H5N1 virus research. The overall country ranking is highlighting the difference in the scientific fields. Especially in the case of H5N1, the origin of the bird-flu was in East-/Southeast Asia. Countries such as Vietnam or Thailand are usually considered to be scientifically marginal with respect to global research performance, while their contribution to H5N1 research is clearly to be seen here. However, highly interesting is that the expected performance of Vietnam was estimated to be even higher as indicated by a very low $p_{rand}$. In contrast, Singapore and Taiwan are more strategically positioned, given the structure of the network. Another spot of significant outbreaks was in Africa (cf. Leung et al. 2011). The involvement of African countries in the global research system on H5N1 is also very pronounced. Egypt takes an exceptional role in knowledge flows, compared to the usual perception of the country in the global science system.

For the nanotube results in general, the simple scale of number of nodes in a group is weakly correlated with either the empirical $C_{RB}$ (r=0.45) or the randomized estimate (r=0.47). The normalized coefficient $p_{rand}$ is uncorrelated with the empirical $C_{RB}$ (r=0.09) and also with the number of nodes (r=0.02). For the H5N1 results, there are weak to moderate correlations between the number of nodes and the empirical / randomized value of $C_{RB}$ (r=0.78/0.42), while the normalized coefficient $p_{rand}$ is almost uncorrelated with either the empirical $C_{RB}$ (r=0.16) and the number of nodes (r=0.04). In summary, the proposed technique reveals novel insight into the complex structures of the performance of territorialized research systems (e. g. countries).

*Time-Series data: Regionalized Networks (ISIVIP-Data)*

This dataset is based on a combination of two sources, one being Thompson Scientific's SCI-Expanded, and one being the Chinese Index Chongqing VIP. The dataset covers biotech related research papers with a Chinese contributor between the years 2003 and 2009 (see Hennemann et al. 2011b for detailed description on the data). In this example, the regionalization was done on the Chinese province level, including Hong Kong and Macao as provinces.

Table 2 shows three time-steps in the evolution of the biotech research network in China. The way in which the network was constructed includes a reasonable amount of foreign organizations. Those are the organizations that co-authored a paper with a Chinese researcher. Due to this fact, the network is virtually surrounded with sparsely cross-connected foreign nodes. This fact influences the behavior of the randomization in a form that it systematically under-estimates the foreign nodes and constantly over-estimates the Chinese nodes (this is why the $p_{rand}$ values are rather high for most provinces). In a way, this is also related to the fact that networks need to be complete to acknowledge the full complexity of the system. However, having that in mind, most results can be interpreted straight-forward.

Table 2: Provincial Performance in Biotech Research in China in the Years 2003, 2006, and 2009 (in descending Order by $C_{RB}$ (emp); Largest Component: a) 431 Nodes /631 Edges, b) 1,187 Nodes / 2,353 Edges, c) 1,962 Nodes / 4,403 Edges) each 500 Configurations/10x Swaps

| | a) 2003 | | | | | b) 2006 | | | | | c) 2009 | | | |
|---|---|---|---|---|---|---|---|---|---|---|---|---|---|---|
| # | Country | No. Nodes | $C_{RB}$ (emp) | $p_{rand}$ | # | Country | No. Nodes | $C_{RB}$ (emp) | $p_{rand}$ | # | Country | No. Nodes | $C_{RB}$ (emp) | $p_{rand}$ |
| 1 | Hong Kong | 9 | 0,044** | 1,361 | 1 | Hong Kong | 13 | 0,022 | 0,970 | 1 | Shaanxi | 47 | 0,011* | 1,125 |
| 2 | Zhejiang | 16 | 0,033* | 1,257 | 2 | Shaanxi | 22 | 0,020** | 1,407 | 2 | Chongqing | 34 | 0,010** | 1,544 |
| 3 | Shanghai | 40 | 0,026* | 1,109 | 3 | Zhejiang | 35 | 0,012** | 1,244 | 3 | Shanghai | 129 | 0,008** | 0,875 |
| 4 | Beijing | 65 | 0,024* | 1,154 | 4 | Shanghai | 75 | 0,012 | 1,035 | 4 | Hong Kong | 18 | 0,008** | 0,615 |
| 5 | Jilin | 9 | 0,023 | 1,038 | 5 | Beijing | 131 | 0,012** | 1,138 | 5 | Beijing | 216 | 0,007* | 1,060 |
| 6 | Shaanxi | 12 | 0,023* | 1,158 | 6 | Sichuan | 26 | 0,010* | 1,243 | 6 | Jiangsu | 97 | 0,006** | 1,196 |
| 7 | Liaoning | 15 | 0,020** | 1,650 | 7 | Jiangsu | 61 | 0,008** | 1,506 | 7 | Zhejiang | 63 | 0,006* | 0,907 |
| 8 | Chongqing | 4 | 0,019 | 0,940 | 8 | Jilin | 30 | 0,007** | 1,421 | 8 | Liaoning | 62 | 0,006** | 1,356 |
| 9 | Guangdong | 27 | 0,018* | 1,266 | 9 | Guangdong | 78 | 0,007 | 1,070 | 9 | Jilin | 31 | 0,005** | 1,228 |
| 10 | Jiangsu | 31 | 0,014** | 1,351 | 10 | Gansu | 26 | 0,007** | 1,523 | 10 | Guangdong | 106 | 0,004* | 0,888 |

Hong Kong is revealed as the most important disseminator of knowledge, but the relative performance is dropping from 2003 to 2009 as indicated by $p_{rand}$, turning Hong Kong's significant over-performance into an expected position. Almost the same is true for Shanghai, while Beijing faces only a moderate decline over time. Some interior provinces like Shaanxi or Chongqing can improve their performance on the basis of their over-expected scores in 2003. This is interesting, because these provinces are generally assessed to play only a minor role in the science and technology system of China. Jiangsu province and the northeastern provinces Liaoning and Jilin are over-performing as well. Zhejiang province, in contrast, is going down straightly especially from 2006 to 2009. The Zhejiang province goes up in efficiency again so that it can be argued that the mediocre performance is just a temporary one.

Over all, both empirical examples deliver straight forward result that are plausible and can be interpreted directly. They can be cross-compared on the basis of their randomized network models. Moreover, the results can be matched to other empirical results, from primary and secondary statistics on the performance of the respective research systems.

However, all methods need to be evaluated with respect to the reliability and validity of the result that they produce. The testing sample is based on the empirical H5N1 network, because estimates on the basis of smaller networks are usually more error-prone and should therefore produce extremer results. The reliability, i.e. how reproducible are the results, can be assessed by the variability of a measurement. The coefficient of variation (CV), defined as the ratio of the standard deviation and the mean, is an indicator for the statistical spread of a probability distribution.

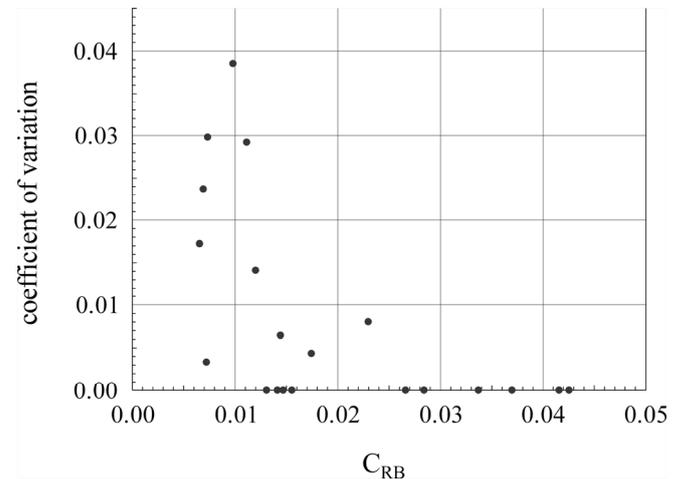

Figure 3: Variation for 100 Repeated Calculations of the randomized Estimates as Function of the empirical Regional Betweenness $C_{RB}$

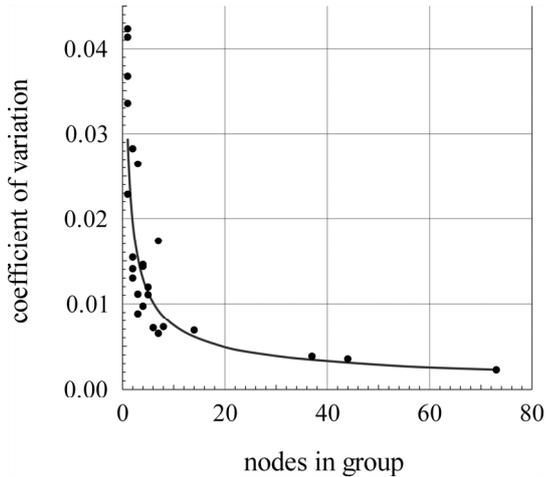

Figure 4: Variation for 100 repeated Calculations of the randomized Estimates as Function of the Number of Nodes in the Groups; the Trend-line suggests a Power-Law with an Exponent of -0.6)

Figure 3 shows the CV as a function of the underlying empirical regional betweenness $C_{RB}$ of the groups. The interrelation between CV and $C_{RB}$ is not very pronounced, but indicates a clear downward trend of the spread with increasing centrality of the group. Figure 4 shows the CV of the estimate as a function of the number of nodes that a group has. Here, the interrelation fit to a power-law with an exponent of -0.6 ($R^2$=0.84). The statistical spread, measured in CV of the estimate drops sharply with increasing group sizes. The simulation of larger networks shows very similar functions (exp -0.7, $R^2$=0.68). For more than 10 nodes in a group, the spread is below 1%, even for networks with several thousand nodes.

Essentially, the results are largely reproducible, independent from the network size and an error is a mainly induced by small absolute number of nodes in a group and the degree of marginality of the empirical position in the network. The more nodes in a group and the more central they are, the smaller is the error.

On the basis of these tests and the experiences build from the work with the empirical networks, some suggestions can be derived to produce reliable estimates:

- Empirical networks should contain at least 100 nodes/100 edges to yield robust estimates.

- The smaller the network, the larger number of repeated edge swaps compared to the empirical number of edges. For large networks, a number of swaps that is 5-10 times higher than the number of empirical networks is sufficient. For small networks with around 100 edges this number should ideally be 20 to 50 times the number of empirical edges.

- The number of samples (configurations) should be higher than 100 for larger networks (250+ nodes), and 200-500 for networks with around 100 nodes.

- No group should contain less than 5 nodes for smaller networks (<1,000 nodes) and no less than 10 nodes for larger networks to keep the results robust. If this is not possible due to empirical data constraints, the results have to be interpreted with caution.

- The more peripheral a group is located empirically the higher is the uncertainty of the estimate.

All these suggestions are influencing the computation time. A valuation of the computing time on an up-to-date desktop computer varies from less than 10 minutes for networks with few hundred nodes and 1,000 edges (500 configurations with 10,000 edge swaps at each step) to more than 5 hours for a network with 4,000 nodes and 18,000 edges (500 configurations with 150,000 edge swaps at each step).

*Discussion*

Many approaches that aim at the evaluation of the regional scientific performance are trying to assess the outcome of scientific activity in terms of citations. This is an approach that induces multiple problems that are related to the qualification of the quantity and the general explanatory power of the concept of citation. Moreover, for newly emerging scientific fields, the number of citations is low, because citing papers take time. Further, the distribution of citations is highly skewed, with only a few papers receive the bulk of citations and most papers receive very little citations.

Here, the evaluation of the regional performance was grounded on a network approach, where co-authorships of scientific papers represent a definable relation between knowledge producing actors in a system. This approach has been followed in the past quite intensively. However, most work is descriptive, although interactions among the system elements are influencing the result, especially in complex systems. The bootstrapping approach in combination with network reshuffling helps to assess the relative performance controlling for the size and density effects of networks. This enables turning descriptive case studies into analytical testing of network positions comparable to inference statistics.

The presented example of the performance evaluation based on network flows can be extended to other indicators, except the degree centrality, which is the preserved property in the null-models. The method is robust, i.e. the results are reproducible and straight-forward in their interpretation. The given empirical examples are just two cases that show the capacity of the approach. However, this concept can just be a first step of methodological development and the application to territorial / group-based network performance. This includes overcoming the shortage of useful data and indicators of science and economic activity.

In this respect, problems with this approach arise from the general difficulty to estimate the error that is incorporated through incomplete data. The central limit theorem is not applying to complex systems, i.e. the completeness of the network remains a critical point. Interestingly, sampling issues are largely ignored in empirical analyses (Newman 2003; Canter and Graf 2008).

Besides the general difficulty with the data, there are mainly three directions into which the research needs to be expanded. First, the presented simplified method itself needs to be benchmarked with other technical approaches that aim the same target, i.e. measuring group performance. This relates to all generic aggregation approaches, e. g. calculating paths followed by straight calculation of the group betweenness, or resilience-based approaches that derive the actual performance relative to a situation of group absence. Further, the test parameter calculation may be compared to a ceteris paribus exponential random graph model (also known as p*). Second, this approach needs to show its robustness for data other than co-authorships. Examples from the global cities interlocking network approach would be of special interest as well as relational datasets that were compiled on the basis of economic collaboration (e.g. co-patenting). Third, the robustness of this approach should be further evaluated through simulation based tests with typical formations of networks. Related to this, the predictive capacity of efficient positions in the past for the future success of a region needs to be assessed. This would mark a turning point in network analysis, because success and failure of nodes/groups of nodes in networks could be explained endogenously.

*Conclusion*

In this article, a novel approach to the evaluation of group performance in science was proposed. It is based on an aggregate of group betweenness (here regions) that is weighted with the overall position in the network measured as k-core number. To be able to assess the empirical results in terms of performance, a randomized network through edge shuffling was used to generate a normalized Null-model. A bootstrapping type re-sampling with subsequent estimation of parameters delivered robust estimates for cross-network comparison.

Two empirical examples were given to enable the reader to assess the validity of the aggregation, the indicator (betweenness), and the estimate. The interpretation of the results for research on H5N1 and nanotubes proved to be plausible. The time-series data on biotechnology related research in China revealed the strength of the method when it comes to evaluate the development of regions.

*Acknowledgements*

Earlier versions of this approach were presented at a workshop on *Structure and Dynamics of Knowledge Networks* (DIME) in 2009 in Eindhoven, at the Network Science Conference 2009 in Venice, and at the *International Conference on Urbanization and Development in China* in Salt Lake City, UT in 2011. The comments given are very much appreciated. I am grateful to Diego Rybski (PIK Potsdam) for fruitful discussions about physicists' methods and sophisticated approaches to networks, and also for pushing me into python.